\documentclass[dvipdfmx]{pasj01}
\usepackage{graphicx}

\begin{document}

\title{The effect of our local motion on the Sandage-Loeb test of
the cosmic expansion}
\author{Takuya \textsc{Inoue}\altaffilmark{1,2,3,*},
Eiichiro \textsc{Komatsu}\altaffilmark{3,4,*},
Wako \textsc{Aoki}\altaffilmark{5,6},
Takeshi \textsc{Chiba}\altaffilmark{7},
Toru \textsc{Misawa}\altaffilmark{8},
Tomonori \textsc{Usuda}\altaffilmark{5,6}}%
\altaffiltext{1}{Department of Mechanical Engineering, Graduate School of Science and Engineering, Doshisha University, 1-3 Tatara Miyakodani, Kyotanabe-shi, Kyoto 610-0394, Japan }
\altaffiltext{2}{Ecole Centrale de Lille, ECLille, Cit\'e Scientifique, F-59650 Villeneuve d'Ascq, France}
\altaffiltext{3}{Max-Planck-Institut f\"{u}r Astrophysik, Karl-Schwarzschild Str. 1, D-85741 Garching, Germany}
\altaffiltext{4}{Kavli Institute for the Physics and Mathematics of the universe, Todai Institutes for Advanced Study, the University of Tokyo,
Kashiwa, 277-8583 Japan(Kavli IPMU, WPI)}
\altaffiltext{5}{National Astronomical Observatory of Japan (NAOJ), 2-21-1 Osawa,
Mitaka, Tokyo 181-8588, Japan}
\altaffiltext{6}{Department of Astronomical Science, School of Physical Sciences, The Graduate University for Advanced Studies, SOKENDAI, 2
-21-1 Osawa, Mitaka, Tokyo 181-8588, Japan}
\altaffiltext{7}{Department of Physics, College of Humanities and Sciences, Nihon University, Tokyo 156-8550, Japan}
\altaffiltext{8}{School of General Education, Shinshu University, 3-1-1 Asahi, Matsumoto, Nagano 390-8621, Japan}
\email{ctwc0518@mail4.doshisha.ac.jp}
\email{komatsu@mpa-garching.mpg.de}


\maketitle

\begin{abstract}
Redshifts of an astronomical body measured at multiple epochs (e.g., separated by 10 years) are different due to the cosmic expansion. This so-called Sandage-Loeb test offers a direct measurement of the expansion rate of the Universe. However, acceleration in the motion of Solar System with respect to the cosmic microwave background also changes redshifts measured at multiple epochs. If not accounted for, it yields a biased cosmological inference. To address this, we calculate the acceleration of Solar System with respect to the Local Group of galaxies to quantify the change in the measured redshift due to local motion. Our study is motivated by the recent determination of the mass of Large Magellanic Cloud (LMC), which indicates a significant fraction of the Milky Way mass.
We find that the acceleration towards the Galactic Center dominates, which gives a redshift change of 7~cm/s in 10 years, while the accelerations due to LMC and M31 cannot be ignored depending on lines of sight. We create all-sky maps of the expected change in redshift and the corresponding uncertainty, which can be used to correct for this effect.
\end{abstract}
\section{Introduction}
Suppose that we measure a redshift of an astronomical body today, and measure it again in 10 years. These redshifts are different due to acceleration or deceleration of the cosmic expansion. Proposed first by \citet{sandage:1962} and elaborated by \citet{loeb:1998}, this effect provides a direct measurement of the expansion rate via \citep[for a review]{Quercellini/etal:2012}
\begin{equation}
\frac{\Delta z}{\Delta t_{0}}= H_{0}(1+z)-H(z)\,,
\end{equation}
where $\Delta z$ is a change in redshift (``redshift drift''), $\Delta t_{0}$ is a time interval (e.g., 10 years), $H(z)$ is the Hubble expansion rate at a given $z$, and $H_0$ is its present-day value. For a flat $\Lambda$ Cold Dark Matter (CDM) model with the matter density parameter of $\Omega_M=0.3$ and $H_0=70~{\rm km/s/Mpc}$, $\Delta z$ is positive at $z\lesssim 2$ and negative otherwise. It is common to express the magnitude of $\Delta z$ in terms of the velocity shift, $\Delta v=c\Delta z/(1+z)$, and we find, for example, $\Delta v\approx -2.5~{\rm cm/s}$ for $z=3$ over $\Delta t_0=10$~years. Measuring this effect is one of the science targets of the upcoming large aperture telescopes such as TMT\footnote{See https://www.tmt.org} and E-ELT \citep{Cristiani/etal:2007,Corasaniti/etal:2007,liske/etal:2008}.

However, our local motion (i.e., the acceleration of Solar System with respect to the rest frame of the cosmic microwave background (CMB)) also contributes to the change in redshift, which will contaminate measurements of the cosmological redshift drift. In this paper we calculate the acceleration of Solar System with respect to the Local Group (LG) of galaxies, to report that $\Delta v$ from the local motion is comparable to the cosmological signal and thus must be corrected. Our study is motivated in particular by the recent determinations of the mass of Large Magellanic Cloud (LMC), which indicate a rather large mass of $(0.15-0.3)\times 10^{12}~M_\odot$ \citep{penarrubia/etal:2016,Laporte/etal:2016,erkal/etal:2019}.

\citet{quercellini/amendola/balbi:2008} and \citet{amendola/balbi/quercellini:2008} considered accelerations of other objects (such as globular clusters) in the Milky Way and member galaxies of galaxy clusters, respectively. We shall not consider them in this paper, but focus entirely on the acceleration of the Solar System with respect to the rest frame of the LG galaxies.

Throughout this paper the error bars are quoted at 68\% confidence level, and the propagation of the errors is done by assuming that the probability distribution is a Gaussian.

\section{Accelerations}
\label{sec:method}
It is convenient to split the local acceleration with respect to the CMB, $\mathbf{a}_{\rm Sun-CMB}$, as
\begin{equation}
\label{eq:acceleration}
   \mathbf{a}_{\rm Sun-CMB} = \mathbf{a}_{\rm Sun-MW} +  \mathbf{a}_{\rm MW-LG}+ \mathbf{a}_{\rm LG-CMB}\,,
\end{equation}
where $\mathbf{a}_{\rm Sun-MW}$, $\mathbf{a}_{\rm MW-LG}$, and $\mathbf{a}_{\rm LG-CMB}$ are the acceleration of Solar System with respect to the center of our Galaxy (Milky Way; MW), that of MW with respect to barycenter of LG, and that of LG with respect to the rest frame of the CMB. However, it is difficult to estimate $\mathbf{a}_{\rm LG-CMB}$ due to the lack of modelling of the structures surrounding LG; thus, we ignore this term and focus on the other two terms in this paper (see, however, the end of Section~\ref{sec:conclusion} for our remark on this term).

In this paper we shall ignore the higher-order (relativistic) corrections, which are at most of order $10^{-3}$ times $\mathbf{a}_{\rm Sun-CMB}$, as they are suppressed by  $v_{\rm Sun-CMB}/c\approx 10^{-3}$. Specifically, we shall include only the acceleration term ($a_r/c$) in Eq.~(A18) of \citet{liske/etal:2008}, or equivalently the 4-acceleration term ($w_{\cal O}^\mu p_\mu$) in Eq.~(3.35) of \citet{korzynski/kopinski:2018}. The general relativistic corrections \citep{korzynski/kopinski:2018,marcori/etal:2018}
are even smaller. See Eq.~(4.16) of \citet{marcori/etal:2018} for the explicit forms of the contribution of the gravitational potentials to the redshift drift. We can estimate the impact of time-derivative of the potential to $\dot z$ as $\dot\Phi\approx H_0(v_{\rm Sun-CMB}/c)^2\approx 10^{-6}~H_0$, which is negligible.

\subsection{Sun-MW}
Let us define our coordinate system such that $x$ is the direction towards the Galactic Center, $y$ is the direction of the Galactic rotation, and $z$ is the direction of the Galactic North Pole. Specifically,
\begin{equation}
\label{eq:coordinates}
x = D\cos{l}\cos{b}\,, \quad
y = D\sin{l}\cos{b}\,, \quad
z = D\sin{b}\,,
\end{equation}
where $l$ and $b$ are the Galactic longitudes and latitudes, while $D$ is the distance to an object. 

First, we calculate $\mathbf{a}_{\rm Sun-MW}$ for a spherically symmetric mass  distribution, and later compare the result with more realistic mass distribution of MW.  By definition, the accelerations in $y$ and $z$ directions vanish. We thus obtain $\ddot{x}=GM(<D)/D^2$, where $G$ is the gravitational constant and $M(<D)$ is the mass of MW inside the orbit of Solar System. Approximating the orbit of Solar System to be circular, we relate $M(<D)$ to the circular velocity $V$ as $M(<D)=DV^2/G$; thus, $\ddot{x}=V^2/D$. Using $D=8.2\pm 0.1~{\rm kpc}$ and $V = 238\pm 15~{\rm km/s}$ \citep{bland-hawthorn/gerhard:2016}, we obtain $\ddot{x}=(2.25\pm0.28)\times10^{-8}~{\rm cm/s^2}$.

We compare this result with that from more realistic mass distribution of MW. Following \citet{kallivayalil/etal:2013} and \citet{gomez/etal:2015}, we model the gravitational potential of MW as the sum of a Miyamoto-Nagai disk \citep{miyamoto/nagai:1975}, a Hernquist bulge \citep{hernquist:1990}, and a Navarro-Frenk-White dark matter halo \citep{navarro/frenk/white:1995} including the effect of adiabatic contraction \citep{gnedin/etal:2004}. See \citet{gomez/etal:2015} for the parameters of the model. The cosmological parameters are $\Omega_m=0.3$ and $H_0=70~{\rm km/s/Mpc}$. Taking a derivative  of  the potential, we find that this model yields the Solar System circular velocity of $239~{\rm km/s}$ and the acceleration of $\ddot{x}=2.27\times10^{-8}~{\rm cm/s^2}$, which are in excellent agreement with those of the spherically symmetric model. Therefore, for simplicity we adopt $\ddot{x}=(2.25\pm0.28)\times10^{-8}~{\rm cm/s^2}$ as our baseline result.

\subsection{MW-LG}
Next, we calculate $\mathbf{a}_{\rm MW-LG}$. As the mass of LG is dominated by MW, M31 and LMC, we approximate the MW-LG dynamics as a three-body problem. Including M33 and SMC, whose masses are one order of magnitude smaller than those of M31 and LMC, respectively, does not change our results significantly. The acceleration of each galaxy is thus given by
\begin{equation}
\ddot{\mathbf{r}}_i =-G\sum_{i\neq j}
M_{j}\frac{(\mathbf{r}_i-\mathbf{r}_j)}{D_{ij}^3}\,,
\end{equation}
where $\mathbf{r}_i=(x_i,y_i,z_i)$ and $D_{ij} = \sqrt{(\mathbf{r}_j-\mathbf{r}_i)\cdot (\mathbf{r}_j-\mathbf{r}_i)}$. Here, $i=$ MW, M31, LMC and $M_i$ is the total mass of each galaxy. We use $(l,b)$ from SIMBAD\footnote{\sf{http://simbad.u-strasbg.fr/simbad/}}, $D$ given in \citet{penarrubia/etal:2016},
and Eq.~(\ref{eq:coordinates}) to calculate the coordinates of galaxies ($x, y, z$). See  Table~\ref{table:positions} for the values. 

\begin{table}
\caption{Coordinates of galaxies}
\label{table:positions}
\begin{center}
\begin{tabular}{lcc}
\hline
 & M31 & LMC \\
\hline
$l$ (deg) & $121.174322$ & $280.4652$\\
$b$ (deg) & $-21.573311$ & $-32.8884$\\
$D$ (kpc) & $783\pm25 $ & $ 51\pm2 $ \\  
\hline
$x$ (kpc) & $-376.9\pm12.0$ & $7.8\pm0.3$\\ 
$y$ (kpc) & $623.0\pm19.9$ & $-42.1\pm1.7$\\ 
$z$ (kpc) & $-287.9\pm9.2$ & $-27.7\pm1.1$\\
\hline
\end{tabular}
\end{center}
\end{table}
\begin{table}
\caption{Mass of galaxies. The error bars have been symmetrized.}
\label{table:mass}
\begin{center}
\begin{tabular}{lc}
\hline
 & Mass ($10^{12}~M_\odot$)\\  
 \hline
 MW & $1.04\pm0.245$ \\
 M31 & $1.33\pm0.36$ \\
 LMC & $0.25\pm0.085$\\ 
\hline
\end{tabular}
\end{center}
\end{table} 

The masses are \begin{math} M_{\rm MW} = (1.04^{+0.26}_{-0.23})\times10^{12}M_\odot \end{math}, \begin{math} M_{\rm M31} = (1.33^{+0.39}_{-0.33})\times10^{12}M_\odot \end{math}, and \begin{math} M_{\rm LMC} = (0.25^{+0.09}_{-0.08})\times10^{12}M_\odot \end{math} \citep{penarrubia/etal:2016}. 
The LMC mass can be slightly higher \citep{Laporte/etal:2016} or lower \citep{erkal/etal:2019} than this value. As we find that the contribution of $\bf{a}_{\rm Sun-MW}$ dominates, the uncertainty in the LMC mass has a minor impact on our result.

\begin{table*}
\caption{Accelerations}
\label{table:acceleration}
\begin{center}
\begin{tabular}{lccc}
\hline
 & ${\bf a}_{\rm Sun-MW}$ & ${\bf a}_{\rm MW-LG}$ & ${\bf a}_{\rm Sun-LG}$ \\
\hline
   $a_x$ (cm/s$^2$) &
   \begin{math}(2.25\pm0.28)\times10^{-8}\end{math} &
   \begin{math}(-2.72\pm1.25)\times10^{-11}\end{math} &
   \begin{math}(2.24\pm0.28)\times10^{-8}\end{math} \\
   $a_y$ (cm/s$^2$) &
   \begin{math}0\end{math} &
   \begin{math}(-1.13\pm0.40)\times10^{-9}\end{math} & 
   \begin{math}(-1.13\pm0.40)\times10^{-9}\end{math}\\
   $a_z$ (cm/s$^2$) &
   \begin{math}0\end{math} &
   \begin{math}(-7.68\pm2.65)\times10^{-10}\end{math} &  
   \begin{math}(-7.68\pm2.65)\times10^{-10}\end{math}\\ 
\hline
\end{tabular}
\end{center}
\end{table*}

For convenience of the error propagation, we symmetrize the error bars and use the values given in Table~\ref{table:mass}.
The estimated accelerations are summarized in Table~\ref{table:acceleration}.

\section{All-sky maps of the velocity shift}
\label{sec:results}
The cosmological redshift drift is of order a few cm/s in 10 years. To compare this, we convert the acceleration to the velocity shift using $\Delta {\bf v}= {\bf a}\Delta t_0$ with $\Delta t_0=10$~years. We find $(\Delta v_x,\Delta v_y,\Delta v_z)=(7.08\pm0.90,-0.36\pm0.13,-0.24\pm0.08)$~cm/s. Therefore, the change in the redshift due to local motion is comparable to the cosmological signal.

Using 30 years of the VLBI data,
\citet{xu/wang/zhao:2012} inferred the acceleration of the Solar System with respect to the rest frame of extra-galactic objects (also see \citet{xu/wang/zhao:2012a} for their earlier estimate). They find $a_x=(7.47\pm 0.46)~{\rm mm/s/yr}$.\footnote{To compare with our estimate in units of ${\rm cm/s}$ over 10 years, we need to divide and multiply their value by 10, which leaves the numerical value unchanged.} The component in the $y$ direction is not detected, $a_y=(0.17\pm 0.57)~{\rm mm/s/yr}$. These results agree with those in the other studies to within  uncertainties \citep{titov/lambert/gontier:2011,truebenbach/darling:2017}. On the other hand, surprisingly, \citet{xu/wang/zhao:2012} find a significant component in the $z$ direction, $a_z=(3.95\pm 0.47)~{\rm mm/s/yr}$, which is not found by the other studies \citep{titov/lambert/gontier:2011,truebenbach/darling:2017}. Our model yields $a_x$ and $a_y$ which are in agreement with all studies to within uncertainties \citep{titov/lambert/gontier:2011,xu/wang/zhao:2012,truebenbach/darling:2017}. While our $a_z$ agrees with \citet{titov/lambert/gontier:2011} and \citet{truebenbach/darling:2017}, it disagrees strongly with \citet{xu/wang/zhao:2012}.

\begin{figure}
 \centering
  %\textbf{\begin{math}\Delta v_{Sun-LG}\end{math}}
 \includegraphics[width=8cm]{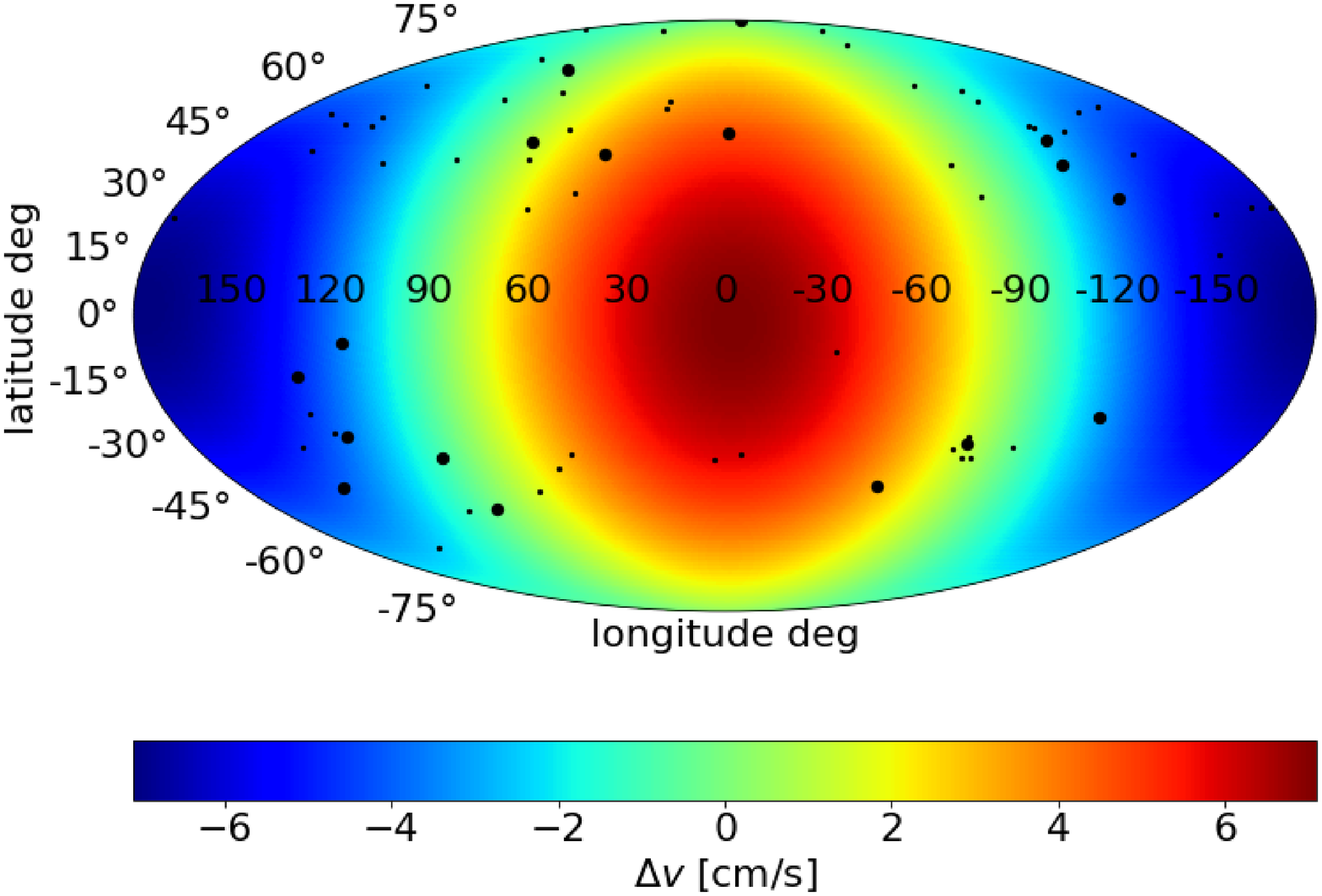}\\
 \includegraphics[width=8cm]{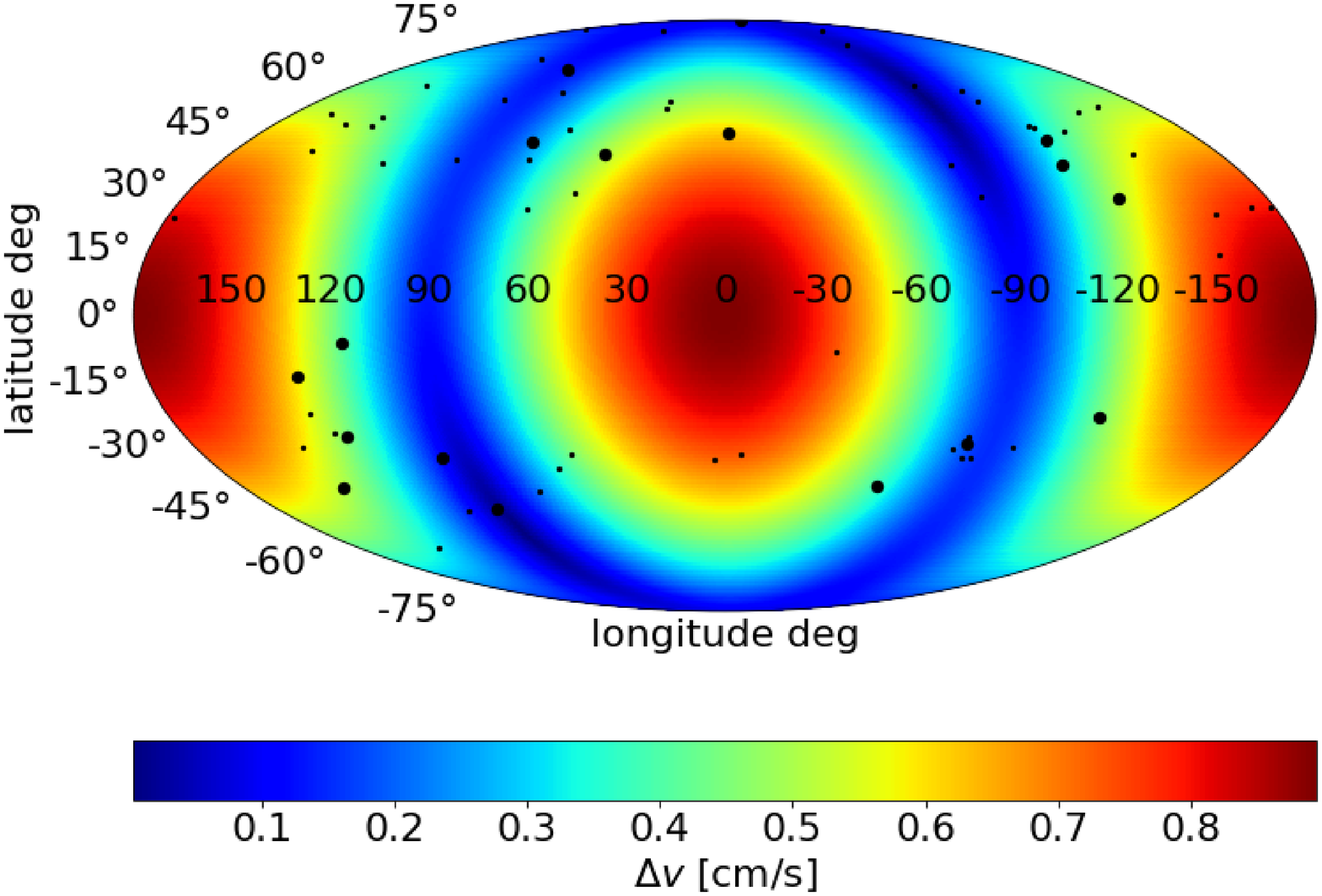}
\caption{All-sky maps of $\Delta v(l,b)$ (top panel) and their uncertainty  (bottom panel) in units of $(\Delta t_0/10~{\rm years})$~cm/s in the Galactic projection. The filled circles show the locations of 71 bright quasars at $z=2-4$ with $B<16.5$ taken from \citet{flesch:2015}, while the bigger circles show those with $B<15.5$.}
\label{fig:Galaxy}
\end{figure}

To create an all-sky map of $\Delta {\bf v}$ in the Galactic coordinates, we use $\Delta v(l,b) = \Delta v_x\cos{l}\cos{b} + \Delta v_y\sin{l}\cos{b} + \Delta v_z\sin{b}$. In the top and bottom panels of Figure~\ref{fig:Galaxy}, we show the maps of $\Delta v(l,b)$ and the uncertainty, respectively, in units of cm/s. We also show locations of 71 bright quasars at $z=2-4$ with the B-band magnitude of $B<16.5$ following  \citet{balbi/quercellini:2007}, taken from the MILLIQUAS quasar catalogue \citep[updated to version 5.7\footnote{\sf http://quasars.org/milliquas.htm}]{flesch:2015}. We find that $\Delta v(l,b)$ is dominated by the acceleration towards the Galactic Center. The maximum value is $7.1$~cm/s. The direction of the maximum, $(l,b)=(-3^\circ,-2^\circ)$, is slightly off the Galactic Center, which indicates importance of the contribution from LG.

\begin{figure*}
  \begin{center}
    \begin{tabular}{c}
      % 1
      \begin{minipage}{0.5\hsize}
      \centering
      \textbf{\begin{math}\Delta v_{Sun-MW}\end{math}(Spherical MW)}
         \includegraphics[width=8cm]{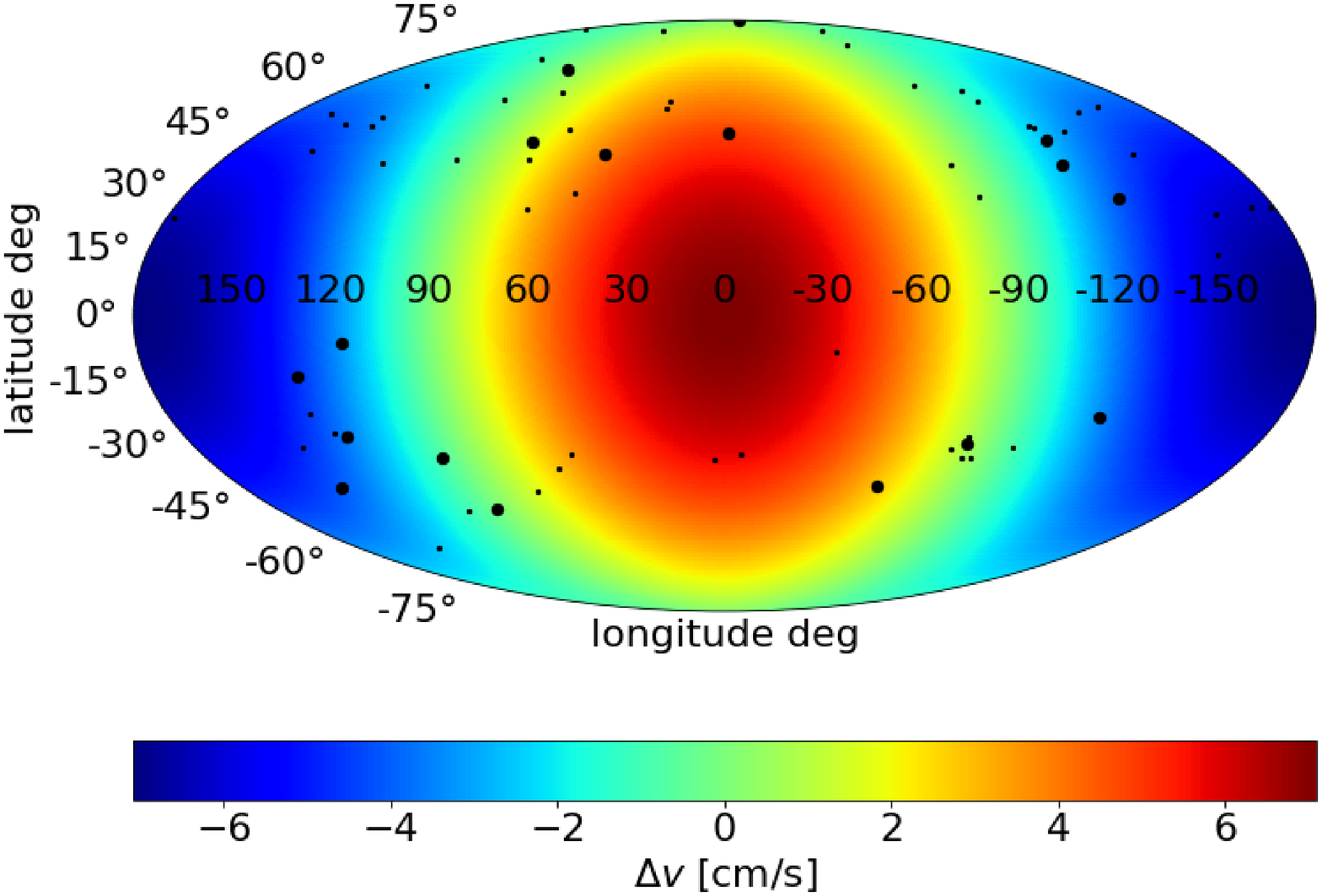}
      \centering
      \textbf{\begin{math}\Delta v_{Sun-MW}\end{math}(Difference between models)}
         \includegraphics[width=8cm]{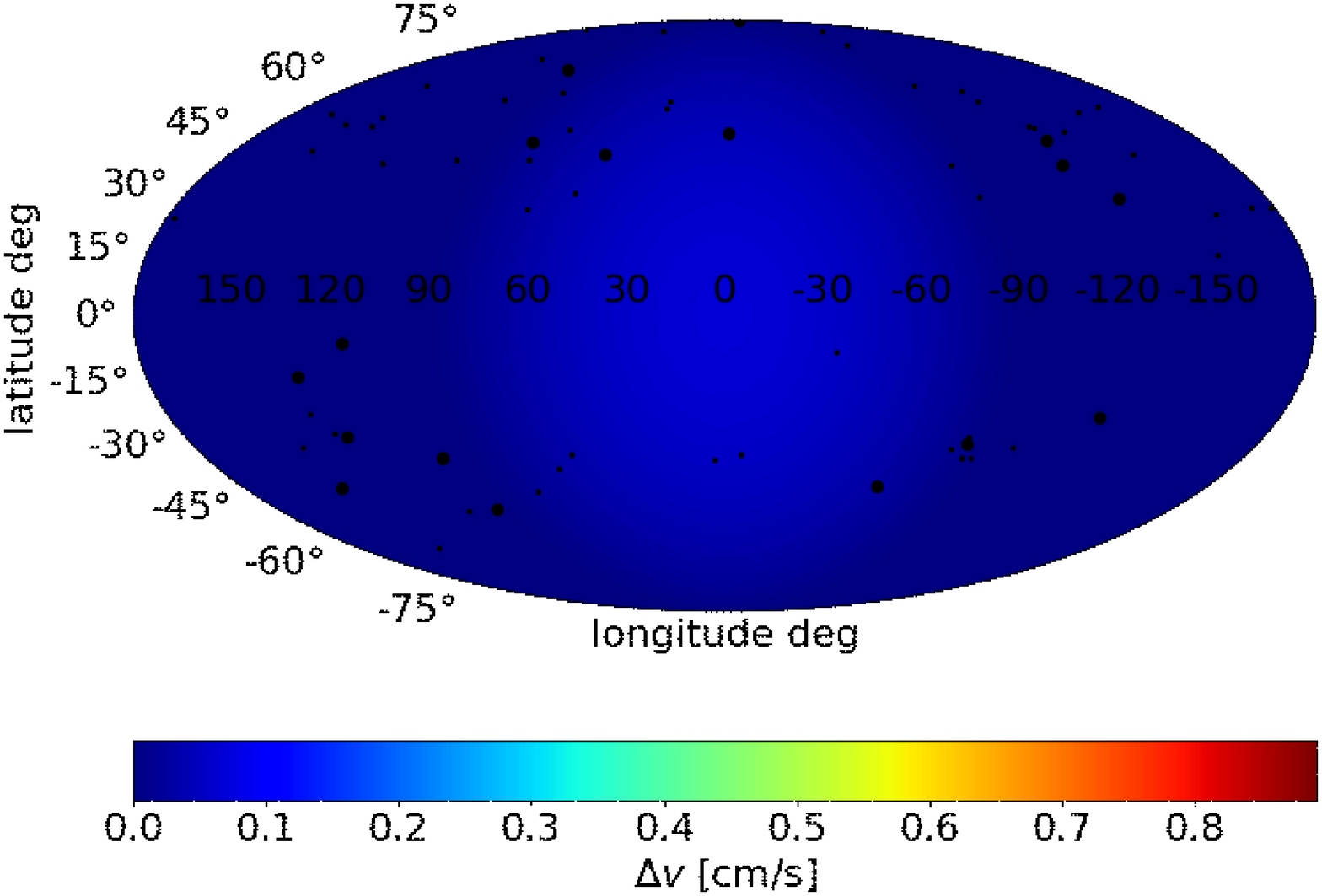}
          \hspace{3cm} 
      \end{minipage}
      % 2
      \begin{minipage}{0.5\hsize}
      \centering
      \textbf{\begin{math}\Delta v_{Sun-MW}\end{math}(Realistic MW)}
          \includegraphics[width=8cm]{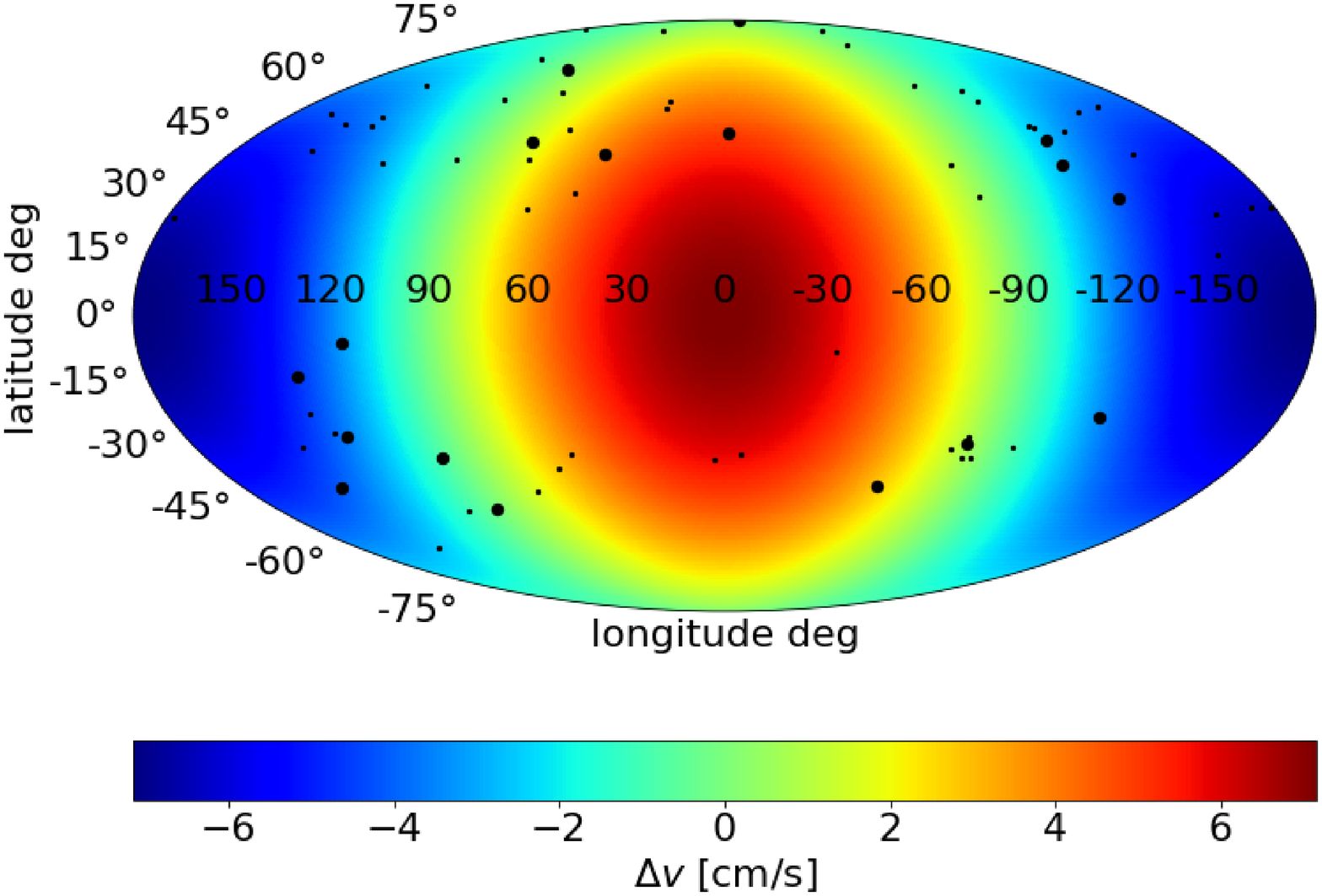}
      \centering
      \textbf{\begin{math}Uncertainty\end{math}}
          \includegraphics[width=8cm]{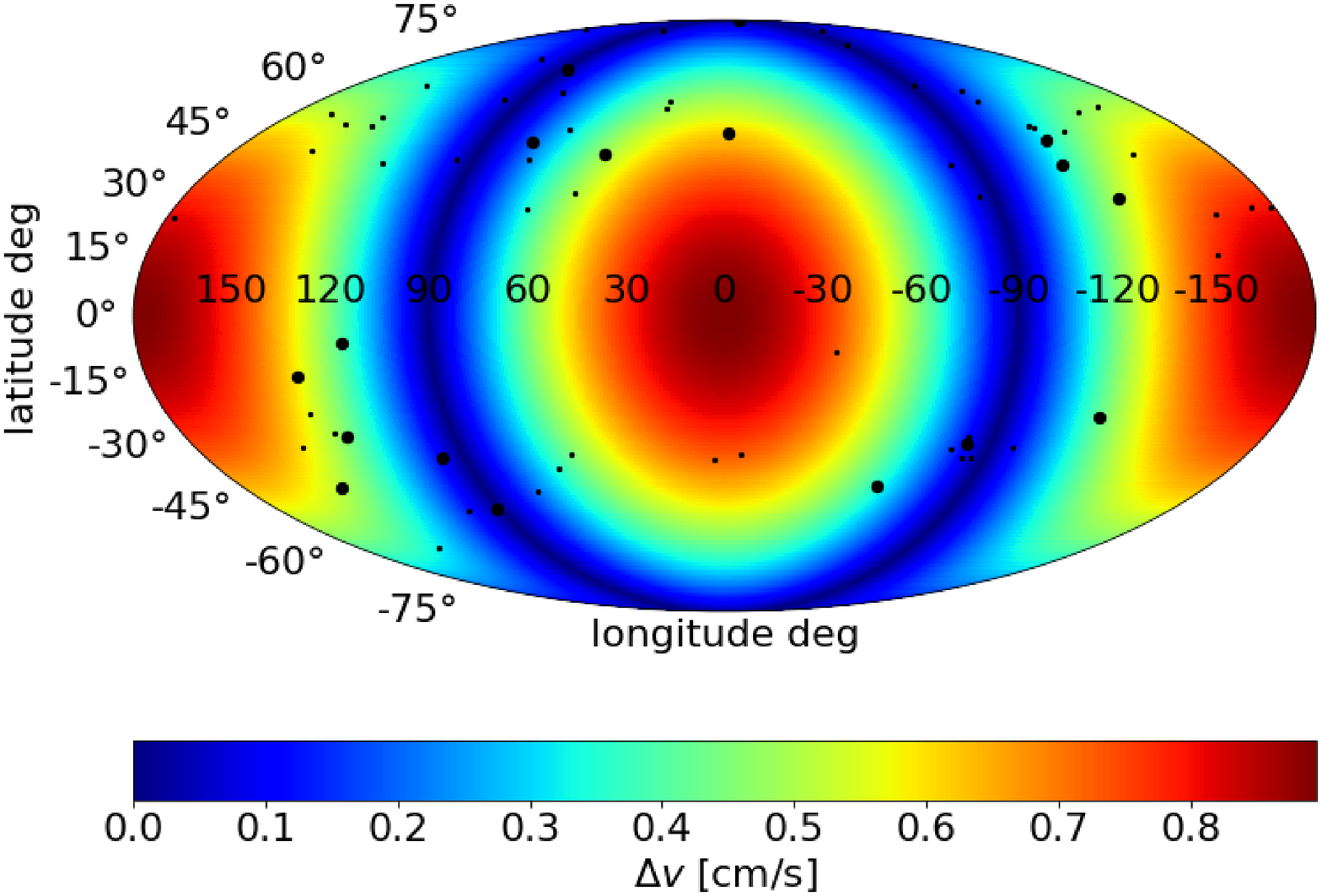}
          \hspace{3cm} 
      \end{minipage}
    \end{tabular}
     \caption{All-sky maps of $\Delta v(l,b)$ from $\bf{a}_{\rm Sun-MW}$ in the spherical model of MW (top left) and more realistic model (top right).
     The bottom left panel shows the difference between these models. The bottom right panel shows the uncertainty of $\Delta v_{\rm Sun-MW}(l,b)$  calculated from  the spherical model.}
    \label{fig:GalaxyComp}
  \end{center}
\end{figure*}

To check accuracy of the results from the spherically symmetric model of MW, we compare $\Delta v_{\rm  Sun-MW}$ from the spherically symmetric model and more realistic model in Figure~\ref{fig:GalaxyComp}. We find that the difference between these two models is below the uncertainty. Note that there is a plane in the uncertainty map where the uncertainty is close to zero. This occurs because this is the direction perpendicular to the Galactic Center, in which the acceleration (hence $\Delta v(l,b)$) vanishes by definition.

\begin{figure*}
  \begin{center}
    \begin{tabular}{c}
      % 1
      \begin{minipage}{0.5\hsize}
      \centering
      \textbf{\begin{math}\Delta v_{Sun-MW}\end{math}}
         \includegraphics[width=8cm]{vsmeannew.eps}
      \centering
         \includegraphics[width=8cm]{deltavsstdnew.eps}
          \hspace{3cm} 
      \end{minipage}
      % 2
      \begin{minipage}{0.5\hsize}
      \centering
      \textbf{\begin{math}\Delta v_{MW-LG}\end{math}}
          \includegraphics[width=8cm]{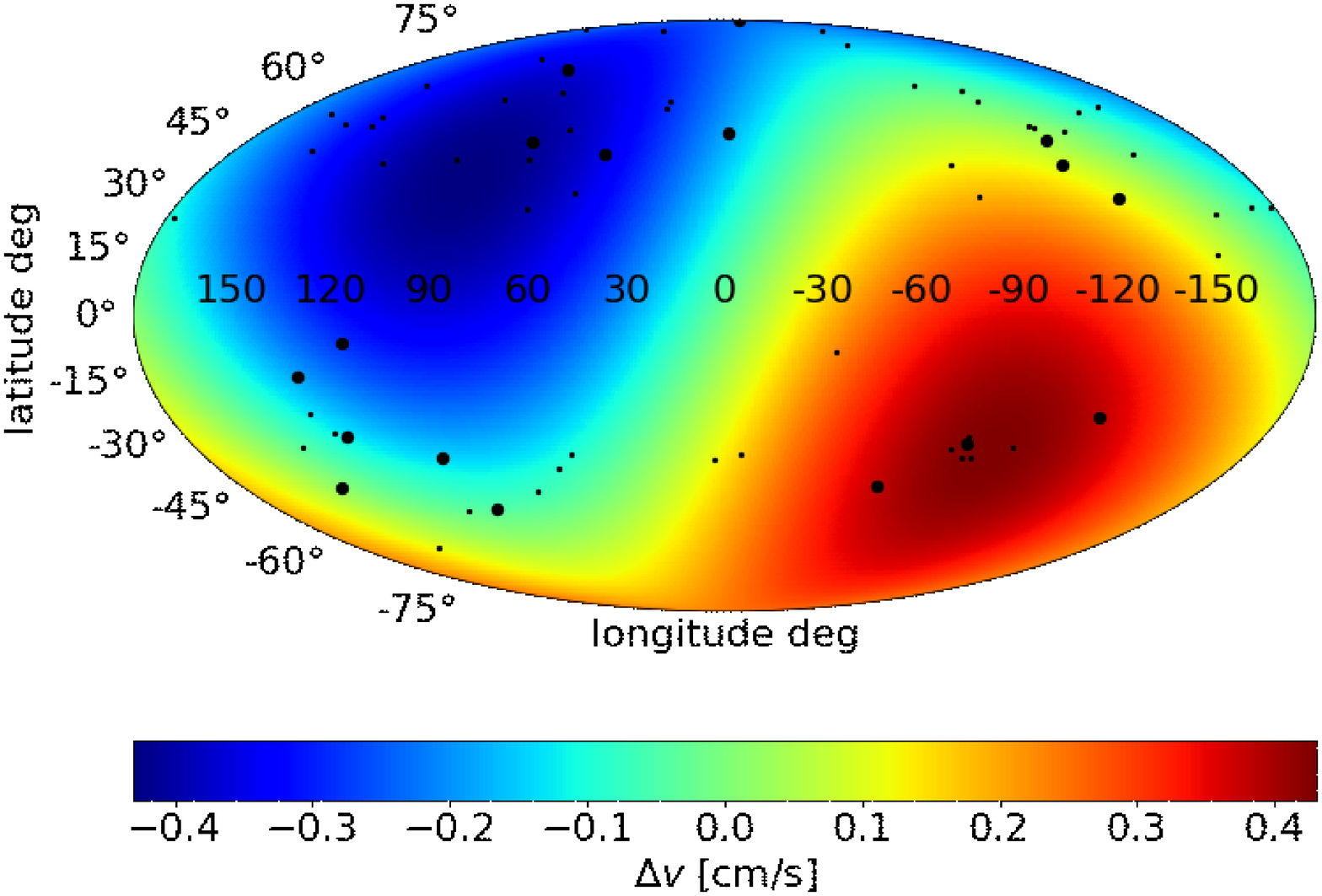}
      \centering
          \includegraphics[width=8cm]{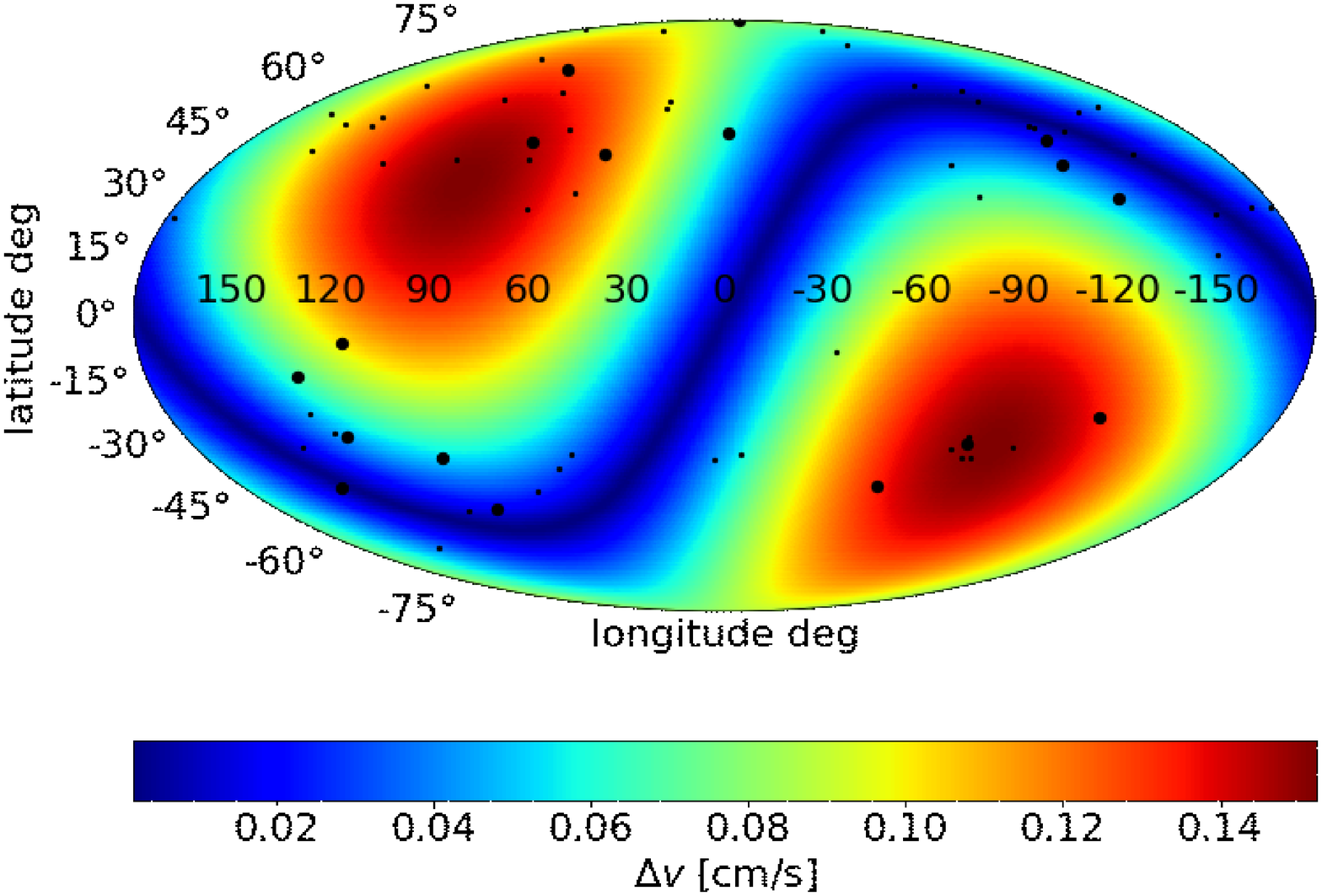}
          \hspace{3cm} 
      \end{minipage}
    \end{tabular}
     \caption{Same as Figure~\ref{fig:Galaxy}, but $\Delta v(l,b)$ from $\bf{a}_{\rm Sun-MW}$ (left) and $\bf{a}_{\rm MW-LG}$ (right).}
    \label{fig:Galaxy2}
  \end{center}
\end{figure*}

To show importance of the contribution from M31 and LMC, we separately show $\Delta v(l,b)$ from $\bf{a}_{\rm Sun-MW}$ (left) and  $\bf{a}_{\rm MW-LG}$ (right) in Figure~\ref{fig:Galaxy2}. We find that the latter is dominated by the acceleration towards LMC, and the maximum value, 0.43~cm/s, is greater than the uncertainty in $\Delta v(l,b)$ in that direction $(l,b)\simeq (280^\circ,-30^\circ)$. Therefore, we cannot ignore the contributions from $\bf{a}_{\rm MW-LG}$.

The contribution of LMC is a factor of a few smaller than typical precision of  redshift drift measurement in the upcoming projects such as TMT and E-ELT, which is given by $\sigma_v=1.35~{\rm cm/s}~({\rm S/N}/2370)^{-1}(N_{\rm QSO}/30)^{-1/2}[(1+z_{\rm QSO})/5]^{-1.7}$ \citep{liske/etal:2008}. Here, ${\rm S/N}$ is the signal-to-noise ratio of spectra per $0.0125$~\AA~pixel, and $N_{\rm  QSO}$ and $z_{\rm QSO}$ are  the number and redshift of quasars, respectively.

\section{Conclusion}
\label{sec:conclusion}
In this paper we have calculated the change in redshift due to the motion of Solar System with respect to LG. We find that the acceleration towards the Galactic Center dominates, which yields a velocity shift of 7.1~cm/s in 10 years. This is comparable to the cosmological effect; thus, we must correct for the velocity shift due to the local motion before inferring the expansion rate of the Universe from redshift drift measurements. 

We also find that the contributions from LMC and M31 (with the maximum velocity shift of 0.43~cm/s towards the LMC direction) 
are greater than the uncertainty in our estimate. They will become more important as we improve upon accuracy of the distance to the Galactic Center and the Galactic rotation velocity. 

While the cosmological redshift drift signal depends on redshifts of objects, the change due to the local motion does not. In addition, the cosmological redshift drift signal is isotropic over the sky, whereas the local motion yields a dipolar pattern. We can also use these properties to subtract the effect of the local motion. When doing so, we must take into account the  fact that quasars at different redshifts appear at different locations in the sky, which receive different contributions  from the local  motion. If we  ignore  this, the local motion would yield a spurious redshift-dependent effect, which could be confused as the cosmological redshift drift.

We have ignored the last term in the left hand side of Eq.~(\ref{eq:acceleration}), $\bf{a}_{\rm LG-CMB}$. If the future measurements of the redshift drift reveal the pattern that is very different from Figure~\ref{fig:Galaxy}, it may indicate that $\bf{a}_{\rm LG-CMB}$ is large. Indeed this might have been seen already from 30 years of the VLBI data indicating a significant vertical acceleration component $a_z$ \citep{xu/wang/zhao:2012}, though other studies \citep{titov/lambert/gontier:2011,truebenbach/darling:2017} do not find such a large $a_z$.

This situation is reminiscent of the dipole anisotropy of the CMB: the direction of dipole was found to be nearly opposite of the direction of the Galactic rotation, which led to unexpected discovery that LG is moving with respect to the rest frame of CMB at a velocity faster than 600~km/s (see Section 4.7.2 of \cite{FBB} for recollection by David Wilkinson). Who knows, a similar surprise might await us.

\begin{ack}
We thank Frank Eisenhauer and Facundo Ariel G\'omez for useful discussion, and anonymous referees for useful comments which improved the paper. T.~I. thanks Kohei Hayashi for useful discussion, 
E.~K. and the Max Planck Institute for Astrophysics (MPA) for hosting his internship, during which this work was initiated and completed, and Benjamin Katryniok and the Ecole Centrale de Lille for giving him the opportunity to do the internship at MPA, as well as for their encouragement and support. This work was supported in part by JSPS KAKENHI Grant Number JP15H05894 and JP15H05896. T.~I. acknowledges a scholarship from Japan Public-Private Partnership Student Study Abroad Program-TOBITATE! Young Ambassador (members from MEXT, JASSO, private sectors and universities). 
\end{ack}

\bibliographystyle{aasjournal}
\bibliography{references}
\end{document}